\documentclass{pnastwo}

\usepackage{cite}
\usepackage{amsmath,amsfonts,bm}
\usepackage{amssymb}
\usepackage{graphicx}
\usepackage{color}

\url{}

\footlineauthor{Katira, et al.}
\volume{}
\issuenumber{}
\issuedate{}

\begin{document}

\title{The order\textendash disorder transition in model lipid bilayers is a first-order hexatic to liquid phase transition}

\author{Shachi Katira\affil{1 $^{\#}$}{Department of Bioengineering, University of California, Berkeley, CA 94720, USA},
Kranthi K. Mandadapu\affil{2 $^{\#}$}{Chemical Sciences Division, Lawrence Berkeley National Laboratory, Berkeley, CA 94720, USA},
Suriyanarayanan Vaikuntanathan\affil{3 $^{\#}$}{Department of Chemistry, University of Chicago, Chicago, IL,  60637, USA},
Berend Smit\affil{4}{Department of Chemical and Biomolecular Engineering, University of California, Berkeley, CA 94720, USA}\affil{5}{Institut des Sciences et Ing\'enierie Chimiques, Valais Ecole Polytechnique F\'ed\'erale de Lausanne Rue de l'Industrie 17, Sion CH-1950 Switzerland}\affil{6}{Department of Chemistry, University of California, Berkeley, Berkeley, CA 94720 USA}
\and
David Chandler\affil{6}{Department of Chemistry, University of California, Berkeley, Berkeley, CA 94720 USA}}

\newcommand\blfootnote[1]{%
  \begingroup
  \renewcommand\thefootnote{}\footnote{#1}%
  \addtocounter{footnote}{-1}%
  \endgroup
}

%\contributor{Submitted to Proceedings of the National Academy of Sciences of the United States of America}

\significancetext{
Lipid bilayers exist in ordered and disordered states --- so-called ``gel'' and ``liquid'' phases.  First-order phase transitions between the two have long been known, but whether the ordered phase is crystal-like or hexatic has not been known.  Here, using large scale molecular simulation, we demonstrate a first-order transition and predict that the ordered phase is hexatic, which can provide a basis for understanding mobility and organization of proteins in the ordered phase.
}

\maketitle

\begin{article}

\begin{abstract} {We characterize the order\textendash disorder transition in a model lipid bilayer using molecular dynamics simulations. We find that the ordered phase appears to be hexatic.  In particular, in-plane structures possess a finite concentration of 5\textendash 7 disclination pairs that diffuse throughout the plane of the bilayer, and further, in-plane structures exhibit quasi-long-range orientational order and short-range translational order. In contrast, the disordered phase is liquid.  The transition between the two phases is first order.  Specifically, it exhibits hysteresis, and coexistence exhibits an interface with capillary scaling.  The location of the interface and its spatial fluctuations are analyzed with a spatial field constructed from a rotational-invariant for local 6-fold orientational order.  As a result of finite interfacial tension, forces of assembly will necessarily exist between membrane-bound solutes that pre-melt the ordered phase.  
}
\end{abstract}


\abbreviations{DPPC (dipalmitoyl phosphatidylcholine) }

\blfootnote{$^{\#}$ SK, KKM, and SV contributed equally to this work. }

In this and the following paper \cite{Katira2015b}, we detail the existence of a first-order order\textendash disorder transition in model lipid bilayers, and we show how pre-melting of the ordered phase is responsible for a powerful membrane-mediated force between transmembrane proteins. Pure lipid bilayers primarily exist in two phases --- an ordered phase often referred to as the `gel' phase (L$_\beta$ or L$_{\beta'}$ phase), and a disordered liquid-like phase (L$_\alpha$ phase) \cite{Tardieu1973}. Here, we use molecular dynamics of the MARTINI model \cite{Marrink2005, Marrink2007} to investigate the nature of the ordered phase and the transition to the disordered phase.

Prior work with the MARTINI model \cite{Marrink2005, Rodgers2012b} on small sections of membrane showed hysteresis when transitioning between ordered and disordered structures.  This form of collective behavior is suggestive of singular behavior, but the 
methods of analysis employed in the earlier works were insufficient to demonstrate the scaling inherent to a first-order transition and whether the ordered phase is a crystal or some other structure. 

In following up on Refs.~\cite{Marrink2005} and \cite{Rodgers2012b}, we establish that a stable interface exists between the ordered and disordered phases at coexistence, and that the fluctuations of the interface are capillary-like \cite{Buff1965, Weeks1977}. These findings establish that the transition between the ordered and disordered membranes is first order, which is consistent with experiments \cite{Mabrey1976}. Further, we show that the spatial structure of the ordered phase exhibits quasi-long-range bond-orientational order and short-range translational order.  These findings indicate that the ordered membrane is a hexatic phase, like those considered in melting of two-dimensional systems~\cite{Nelson, Halperin1978, Marcus1997}, but experiments have not yet determined whether ordered phases of membranes have this structure. 

A molecular model that reproduces the phase behavior of a lipid bilayer is needed to understand lipid behavior around solutes such as proteins and cholesterol, and therefore to understand lipid-mediated interactions \cite{Katira2015b, Machta2012}.  While atomistic-level models \cite{Berger1997,Klauda2010} provide fine-scale structural details, they are far too computationally expensive to study the phase behavior of a bilayer system with any reasonable degree of rigor. To enable access to sufficiently large length scales and time scales (admittedly at the cost of reduced detail), coarse-grained models have been constructed for lipid bilayers and studied using molecular simulations \cite{Marrink2005, Marrink2007, Cooke2005, Venturoli2005, Lenz2005}.  Some of these models exhibit a possibly solid-like ordered phase, and a liquid-like disordered phase. 

We have chosen the MARTINI model because it successfully reproduces equilibrium bilayer properties such as the area per lipid and bending modulus \cite{Marrink2007}. Additionally it has been applied to studies of the plasma membrane \cite{Ingolfsson2014}, lipid rafts \cite{Risselada2008}, organization of transmembrane proteins \cite{Schafer2011}, and membrane tethers \cite{Baoukina2012}. Although we choose dipalmitoyl phosphatidylcholine (DPPC) as a model lipid, we expect our results to hold for many other lipid bilayers or monolayers that exhibit ordered and disordered phases.

\subsection{Order\textendash disorder transition in a model lipid bilayer}

\begin{figure}[t]
\centering
\includegraphics[scale=0.4]{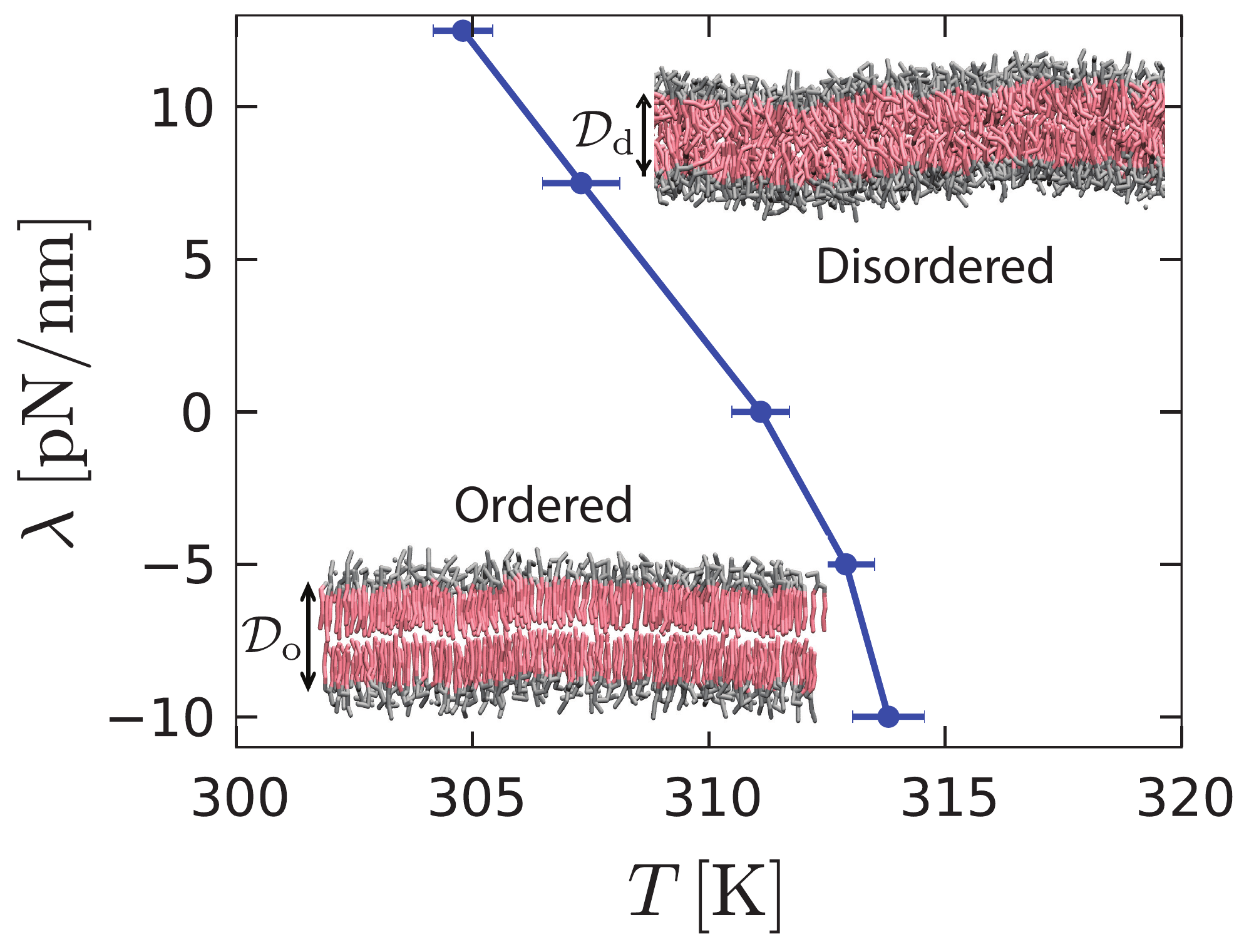}
\caption{\label{fig:PhaseDiag} Order--disorder phase diagram in the tension--temperature, $\lambda$--$T$, plane.  The lateral pressure across the membrane is $-\lambda$.  Points are estimated from 10 independent heating runs like those illustrated in Fig.~\ref{fig:hysteresis} for a periodic system with 128 lipids. Insets are cross sections showing configurations of a bilayer with 3200 lipids in the ordered and disordered phases. The heads are colored gray while the tails are colored pink. Water particles are omitted for clarity.  The hydrophobic thicknesses, ${\mathcal{D}}_\mathrm{o}$ and ${\mathcal{D}}_\mathrm{d}$, are the average vertical distances from the first tail particle of the upper monolayer to that of the lower monolayer.   A macroscopic membrane buckles for all $\lambda < 0$.}
\end{figure}

Fig.~\ref{fig:PhaseDiag} contrasts configurations and shows our estimated phase boundary between ordered and disordered phases in the DPPC MARTINI bilayer system.  The ordered phase has regular tail packing compared to the disorganized tail arrangement of the disordered phase. A consequence of the regular tail packing is that hydrophobic thickness of the ordered phase, ${\mathcal{D}}_\mathrm{o}$, is larger than that of the disordered phase, ${\mathcal{D}}_\mathrm{d}$.  Correspondingly, the area per lipid in the ordered phase is smaller than that in the disordered phase.  

\begin{figure*}[t]
\centering
\includegraphics[scale=0.4]{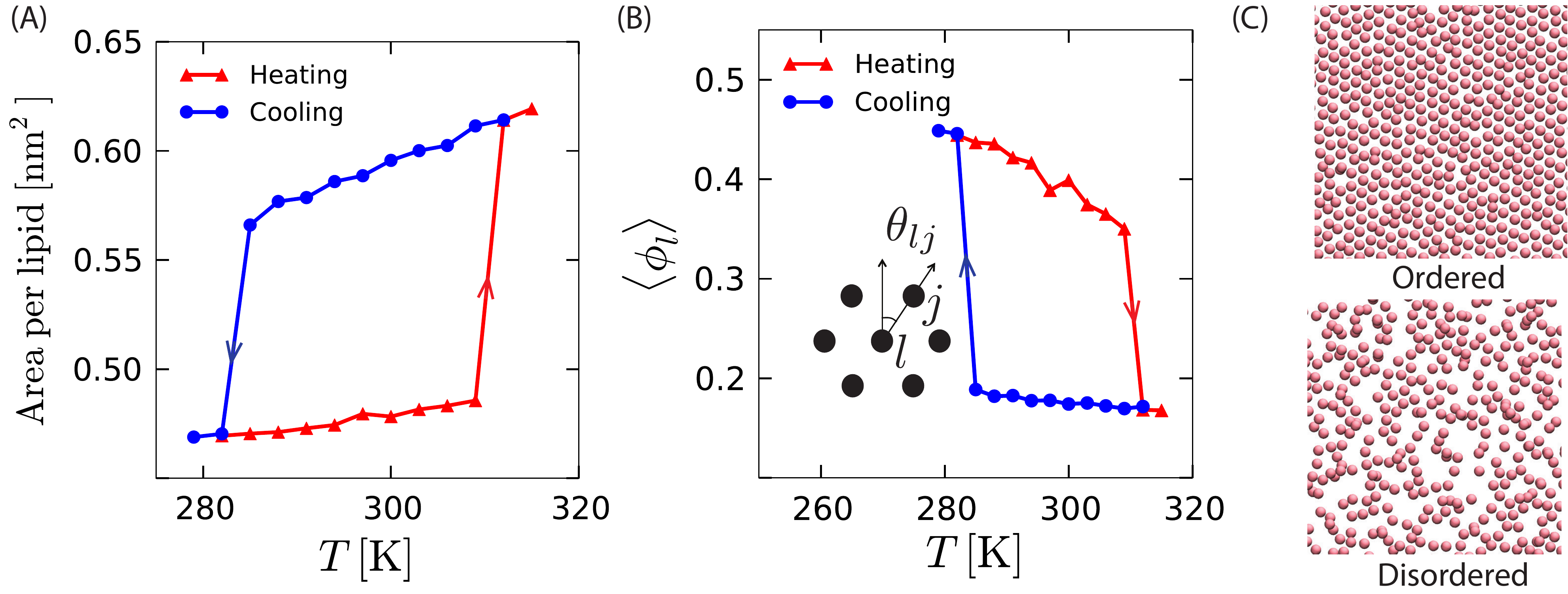}
\caption{Structural measures of different phases as a function of temperature, $T$. (A) Variation in area per lipid with temperature during heating and cooling shows finite jumps and hysteresis. (B) Average local orientational order, $\langle\phi_l\rangle$, also shows finite jumps as a function of temperature while heating and cooling. Magnitudes of heating and cooling rates are 3 K/$\mu$s. (C) Snapshots from the ordered phase at 279\,K (top) and the disordered phase at 315\,K (bottom).  Both display the tail-ends of each lipid in one monolayer of the bilayer. The ordered phase exhibits a hexagonal packing, while the disordered phase exhibits a random arrangement of the tail-end particles.  Empty regions in the disordered configuration are mostly filled by tail-ends from the other monolayer or by non tail-end particles, none of which are not shown in the snapshot.}
\label{fig:hysteresis}
\end{figure*}

The tension\textendash temperature, $\lambda$\textendash $T$, phase diagram shown in Fig.~\ref{fig:PhaseDiag} was estimated from heating runs of a small system --- 128 lipids solvated by 2000 water particles. Fig.~\ref{fig:hysteresis}A shows the change in area per lipid with temperature while heating and cooling a bilayer. There are finite jumps in area per lipid as the system transitions between the two phases, suggesting a first-order phase transition. Hysteresis occurs because ordering from the metastable disordered phase is much slower than disordering from the metastable ordered phase.  Due to the difference in time scales, when contrasting melting and freezing from heating and cooling runs, the melting points from heating runs provide the more accurate estimates of the actual phase boundaries.  The phase boundary graphed in Fig.~\ref{fig:PhaseDiag} shows error estimates  based upon the statistics of several heating runs.  Systematic errors due to small system size and heating rate have not been estimated.

Viewing the end particles of all the lipid chains in one of the two monolayers provides a convenient visual representation that distinguishes the two phases. (Which of the two monolayers chosen for this rendering is irrelevant.)  This representation is shown in Fig.~\ref{fig:hysteresis}C for the ordered and disordered phases. These tail-end particles appear hexagonally-packed in the ordered phase and randomly arranged in the disordered phase. 

To quantify this impression, we consider Halperin and Nelson's local rotational-invariant \cite{Nelson,Halperin1978},
\begin{equation}	
\label{eq:density}
\phi_l=\Big|\,\frac{1}{6}\sum_{j \in \mathrm{nn}(l)} \exp(6\,i\,\theta_{lj})\,\Big|^2\, , 
\end{equation}	
where $\theta_{lj}$ is the angle between an arbitrary axis and a vector connecting tail-end particle $l$ to tail-end particle $j$. Here, the summation over $j\in \mathrm{nn}(l)$ is over the six nearest neighbors of particle $l$. (See inset in Fig.~\ref{fig:hysteresis}B.) The ensemble average, $\langle\phi_l\rangle$, is unity for a perfect hexagonal packing, and it is zero to the extent that hexagonal packing is entirely absent. The variation in $\langle \phi_l \rangle$ with temperature for a DPPC bilayer is shown in Fig.~\ref{fig:hysteresis}B. The finite jumps in $\langle \phi_l \rangle$ as a function of temperature, and the hysteresis, are again suggestive of a first-order phase transition.

\subsection{Stable interface between the ordered and disordered phases exhibits capillary scaling}
\begin{figure}
\centering
\includegraphics[scale=0.4]{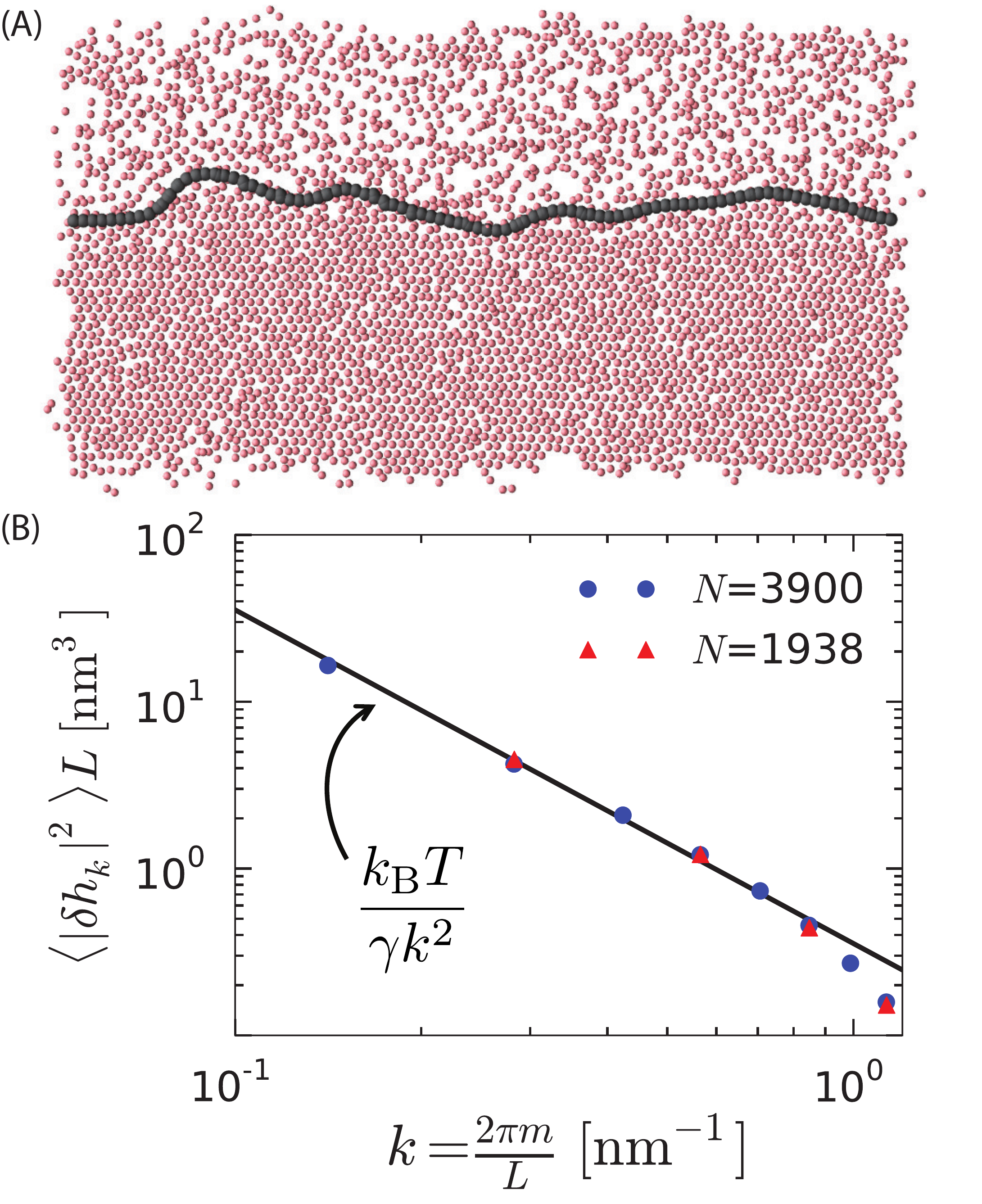}
\caption{Capillary fluctuations of the order\textendash disorder interface: (A) Snapshot of a system showing coexistence between the ordered and disordered phases. The gray contour line indicates the location of the interface separating the ordered and disordered regions. The snapshot is a top view of the bilayer showing the tail-end particles of each lipid in one monolayer. (B) Fourier spectrum of the fluctuations of the instantaneous order\textendash disorder interface. The line is the expected small-$k$ scaling from capillarity theory.\label{fig:FreeInterface}}
\end{figure}

To establish whether the first-order-like behavior described above persists to larger scales and thus actually manifests a phase transition, we consider larger systems and the behavior of the interface that separates the ordered and disordered phases. Fig.~\ref{fig:FreeInterface}A shows this coexistence for a system size of $N=3900$ lipids with an interface between the two phases.  The interface, which spans the membrane, is equilibrated in the constant area ensemble.  This ensemble can maintain an area per lipid that is intermediate between two phases and can therefore stabilize an interface if, in fact, two distinct phases do exist. At such conditions, a line tension can then be calculated from the power spectrum of the interfacial fluctuations. 

To analyze interfacial fluctuations, we first identify the location of the interface at each instant.  This location is found with a two-dimensional version of the three-dimensional construction described in Refs.~\cite{Willard2010,Limmer2014}. Specifically, the interface is here defined as a line in the plane of the bilayer where the value of a coarse-grained field characterizing local orientational order is intermediate between those of the ordered and disordered phases.  The field we use is
\begin{equation} \label{eq:coarsegrainedfield}
\bar{\phi}(\mathbf{r},t) = \sum_l f(\mathbf{r}-\mathbf{r}_l ; \xi)\, \phi_l\, ,
\end{equation}
where $\mathbf{r}_l$ is the position of the $l$th tail-end particle projected onto a plane parallel to that of the bilayer, 
and $f(\mathbf{r}-\mathbf{r}_l ,\xi)$ is a coarse-grained delta-like function in the two-dimensional space,
\begin{equation}
\label{eq:cgGaussian}
f(\mathbf{r}-\mathbf{r}_l ;\xi) = (1/2 \pi \xi^2) \, \exp\left(-|\mathbf{r}-\mathbf{r}_l |^2/2 \xi^2\right) \, .
\end{equation}
Again, the tail-end particles are from the chains of only one of the two monolayers forming the bilayer.  The field variable, $\mathbf{r}$, is a two-dimensional vector specifying a position in the plane of the bilayer. The coarse-graining width, $\xi$, is chosen to be the average separation between tail-end particles $l$ and $j$ when $\langle(\phi_l-\langle\phi_l\rangle)~(\phi_j-\langle\phi_j\rangle)\rangle/\langle(\phi_l-\langle\phi_l\rangle)^2\rangle$ in the ordered phase is 1/10. This choice yields a value of $\xi=1.5\, \mathrm{nm}$. 

For numerics, a square lattice tiles the average plane of the bilayer, and the coarse-grained field $\bar{\phi}(\mathbf{r})$ is evaluated at each lattice node. For convenience, the coarse-graining function is truncated and shifted to zero at $3\xi$. 
The instantaneous order\textendash disorder interface is identified by interpolating between these adjacent lattice nodes to find the set of points $\mathbf{s}$ satisfying $\bar{\phi}(\mathbf{s},t) = (\phi_\mathrm{d}+\phi_\mathrm{o})/{2}$. Here $\phi_\mathrm{d}$ and $\phi_\mathrm{o}$ are $\langle \bar{\phi}(\mathbf{r}) \rangle$ evaluated in the disordered and ordered phases, respectively.

Fig.~\ref{fig:FreeInterface}A shows a snapshot of the instantaneous interface identified in this way. This free interface is stable for the length of simulations, $1.8\, \mathrm{\mu s}$.  For the thermodynamic state considered in that case, zero lateral pressure and 294\,K, we have $\phi_\mathrm{d} = 0.4\pm 0.02$\,nm$^{-2}$ and $\phi_\mathrm{o}=2.15 \pm 0.2$\,nm$^{-2}$.  

Fig.~\ref{fig:FreeInterface}B shows the Fourier spectrum of the height fluctuations of this interface, $\langle |\delta h_k |^2 \rangle$. Two different system sizes are studied, with the larger system having approximately double the interface length of the smaller system. The Fourier component $\delta h_k$ is related to the height fluctuation $\delta h(x)$ as $\delta h(x) = \sum_k \delta h_k \exp (ikx)$ where $x$ is a point on the horizontal axis in Fig.~\ref{fig:FreeInterface}A. Here, $0\leqslant x \leqslant L$, and $L$ is the box size. With periodic boundary conditions, $k=2\pi m/L$, $m = 0,\pm 1, \pm 2, \cdots$. According to capillarity theory \cite{Weeks1977, Safran1994}, $\langle |\delta h_k |^2 \rangle \sim {k_\mathrm{B} T}/{L\gamma k^2}$ for small $k$, with $k_\mathrm{B}$ being the Boltzmann's constant. Given the proportionality with $1/k^2$ at small $k$ (i.e., wavelengths larger than 10\,nm), comparison of the proportionality constants from simulation and capillarity theory determines the interfacial line tension, yielding $\gamma = 11.5\pm 0.46\, \mathrm{pN}$. (This value is significantly larger than the prior estimate of line tension for this model, $3\pm 2\, \mathrm{pN}$~\cite{Marrink2005}.  That prior estimate was obtained from simulations of coarsening of the ordered phase.)

The stability of the interface and the quantitative consistency with capillary scaling provide our evidence for the order\textendash disorder transition being a first-order transition in the model we have simulated.

\subsection{The ordered phase is a hexatic phase}
\begin{figure*}
\centering
\includegraphics[scale=0.4]{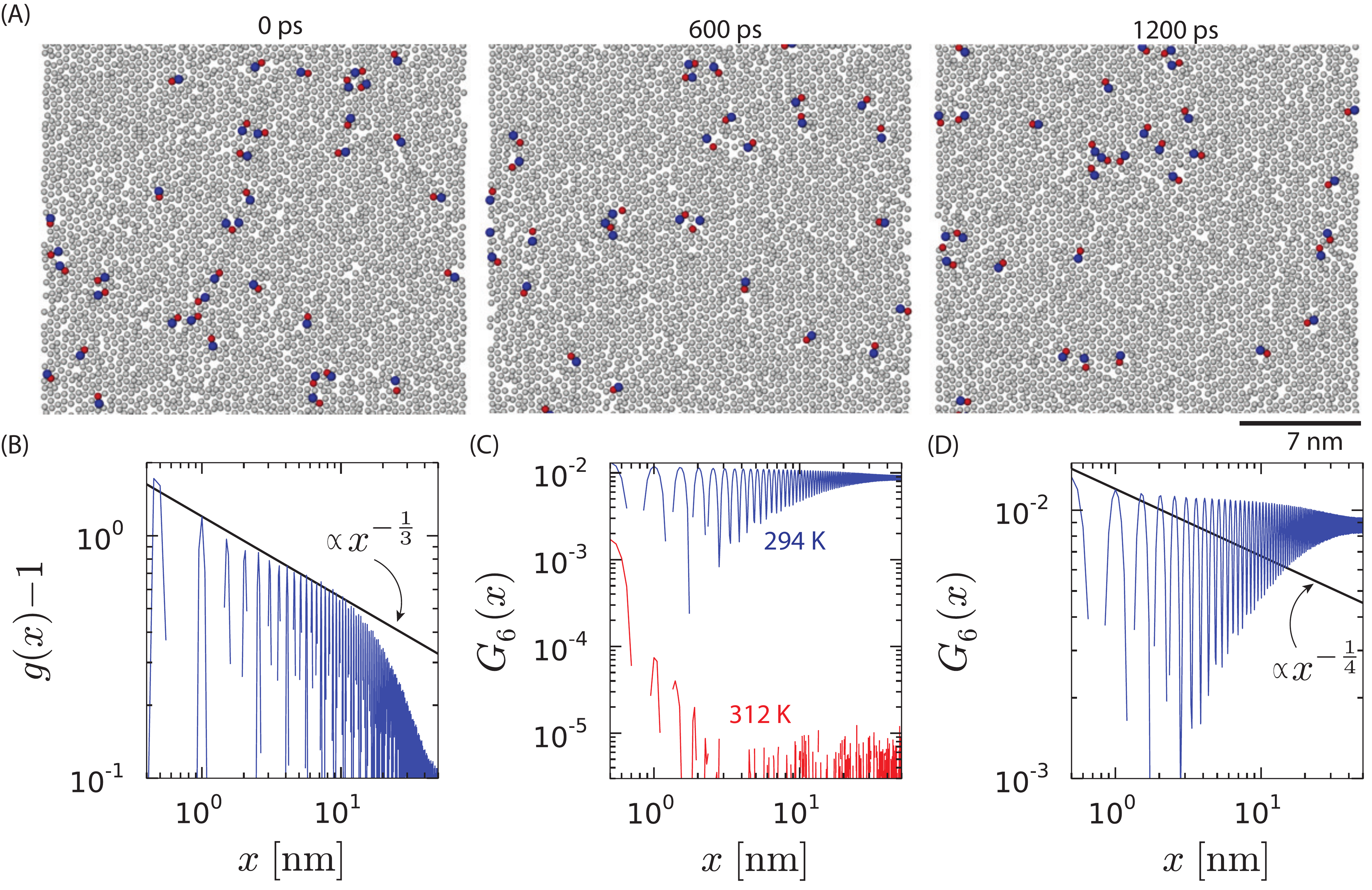}
\caption{\label{fig:correlationFunctions}Hexatic phase: (A) Dislocations (5\textendash 7 disclination pairs) in a 25$\times$25\,nm$^2$ patch of bilayer taken from a 107$\times$110\,nm$^2$ sample in the ordered phase at 294\,K. The tail-end particles of one monolayer are shown in gray. Particles with seven neighbors are highlighted in red, while particles with five neighbors are highlighted in blue. There exist several unbound dislocations as expected for a hexatic phase. These dislocations appear to diffuse freely throughout the system as seen from three different configurations separated by 600\,ps. (B) One-dimensional pair correlation function $g(x)$ showing decay faster than $x^{-1/3}$. (C) Orientational correlation function $G_6(x)$ showing quasi-long-range correlations at $T=294$\,K and short-range correlations at $T=312$\,K.  Values of $G_6(x)$ smaller than $10^{-5}$ amount to statistical noise for the system sizes considered. (D) Enlargement of Panel C to show the slow decay of the quasi-long-range orientational correlations at $T=294$\,K.}
\end{figure*}

The ordered phase shows hexagonal packing with a large number of unbound dislocations, each composed of 5\textendash 7 disclination pairs \cite{Nelson}. These pairs are shown in Fig.~\ref{fig:correlationFunctions}A, which renders a $25\times25\, \mathrm{nm^2}$ bilayer in the ordered phase simulated at $294 \, \mathrm{K}$ and zero lateral pressure. By showing three different configurations of the bilayer separated by $600\, \mathrm{ps}$, we illustrate how the disclination pairs diffuse freely throughout the system. Given that a lipid bilayer is a quasi-two-dimensional system, it is possible that the ordered phase with its unbound dislocations could be in the so-called hexatic phase, an intermediate phase between the solid and liquid phases in two dimensional systems \cite{Nelson, Halperin1978, Nelson1982, Kosterlitz1973, Kosterlitz1978, Nelson1979, Young1979}. The hexatic phase was first predicted for a purely two-dimensional system~\cite{Halperin1978}, and it has been predicted to occur in three-dimensional systems~\cite{Birgeneau1978}.  It has been observed in three-dimensional systems, e.g., Ref.~\cite{Pindak1981}, and in quasi-two-dimensional systems, e.g., Refs.~\cite{Marcus1997, Kjaer1987, Brock1989, Zangi1998}.  This phase is characterized by short-range translational order and quasi-long-range orientational order.

To test whether the ordered phase of the chosen model lipid bilayer system is hexatic, we calculate appropriate translational and orientational correlation functions.  Because any finite sample is anisotropic in the presence of quasi-long-range order, the correlation functions must be applicable when different system orientations are not equivalent.  At the same time, they cannot be based upon a physical reference lattice because the system, we shall see, lacks long-range translational order and is thus not crystalline.  Bernard and Krauth~\cite{Bernard2011} showed how to proceed under such circumstances, and we follow their approach.  

To do so for our membrane system, we again project positions of tail-end particles from one of the monolayers onto the average plane of the bilayer.  Translational order is then examined by considering the pair correlation function along a specific sample orientation. The pair correlation function is 
\begin{equation}
\label{eq:gx}
g(\mathbf{r}) = \frac {1}{\rho}  {\Big\langle \frac{1}{N} \sum_{l(\neq j)} \delta (\mathbf{r} - \mathbf{r}_l + \mathbf{r}_j) \Big\rangle}     \, ,
\end{equation}
where $\delta(\mathbf{r})$ is the delta function for $\mathbf{r}$ in the plane of the bilayer, and $\rho$ is the number of particles per unit area.  Its evaluation along a specific direction, $g(x)$, is $g(\mathbf{r})$ with $\mathbf{r} =x \,\hat{\mathbf{x}}$, where $\hat{\mathbf{x}}$ is the unit vector of an axis of the Cartesian coordinate system for which $\Psi=(1/N)\sum_{l} \Psi_l$ is maximal. Here, $\Psi_l = (1/6)\sum_{j\in\mathrm{nn}(l)}\exp(6\,i\,\theta_{jl})$, with the summation taken over the six nearest neighbors $j$ of particle $l$. 
This pair correlation function, $g(x)$, is denoted by $g(x,0)$ in Ref.~\cite{Bernard2011}.   

For a solid with long-range translational order, $g(x)$ will exhibit a power law decay, specifically $g(x)-1 \sim x^{-\eta}$, with $\eta \le 1/3$ \cite{Strandburg1988},  but our results in Fig.~\ref{fig:correlationFunctions}B show a more rapid decay, seemingly exponential.  This rapid decay is indicative of short-range translational order characteristic of a hexatic phase \cite{Nelson}.  
%
%It has the same asymptotic decay as another correlation function defined by Halperin and Nelson \cite{Nelson}.  Comparison between $g(x)$ and the Halperin--Nelson correlation function is given in the Supporting Information (\textit{SI}).  Unlike the Halperin--Nelson function, $g(x)$ does not require an arbitrary choice of reciprocal lattice vectors.  For a solid with long-range translational order, either of these functions will exhibit a power law decay, but our results in Fig.~\ref{fig:correlationFunctions}B show a more rapid decay, indicative of short-range translational order characteristic of a hexatic phase \cite{Nelson}. 
Our calculations for the ordered phase shown in Fig.~\ref{fig:correlationFunctions} are for a system of 51,200 lipids (solvated by 800,000 water particles) at temperatures $T=294$\,K.  The initial configuration was created by periodically replicating a smaller system.  By construction, therefore, the initial configuration is effectively crystalline, with a large unit cell (or tile).  The equilibrated system with short-range translational order emerges only after several microseconds of simulation.  See Methods and Supporting Information (\textit{SI}).  
%A decay proportional to $x^{-1/3}$ is the theoretical boundary below which a solid phase is no longer stable \cite{Strandburg1988}. The decay of our computed $g(x)$ is below that boundary for the system size we have considered.  

For orientational correlations, we consider
\begin{equation}
\label{eq:psi6}
\psi_6(\mathbf{r})=\frac{1}{6}\sum_{l} \sum_{j\in\mathrm{nn}(l)} \exp (6\,i\,\theta_{lj}) \, \delta(\mathbf{r}-\mathbf{r}_l)\, .
\end{equation}
and compute
\begin{equation}
\label{eq:g6}
G_6(\mathbf{r})= \rho^{-2}\, \langle \psi_6(\mathbf{r}) \, \psi_6^\ast(\mathbf{0}) \rangle\,.
\end{equation}
We then determine $G_6(x)$, which is $G_6(\mathbf{r})$ evaluated at $\mathbf{r} = x\,\hat{\mathbf{x}}$, where again $\hat{\mathbf{x}}$ is the unit vector along an axis of the Cartesian coordinate system that maximizes $\Psi$.  For the hexatic phase, this correlation function will be quasi-long-range (power-law decay), progressing to zero more slowly than $x^{-1/4}$~\cite{Nelson1982, Strandburg1988}. Our results, shown in  Fig.~\ref{fig:correlationFunctions}C and D, are consistent with that behavior.
%, seemingly progressing to zero with a power law much weaker than $x^{-1/4}$. (The theoretical boundary between liquid and hexatic is a decay $x^{-1/4}$ \cite{Strandburg1988}.) 
While the system size is large, it is not large enough to determine the exponent for $G_6(x)$'s decay.  Nevertheless, it is clear from the figure that orientational correlations in the ordered phase are not long ranged, but rather quasi-long ranged over the distance scales accessible to our simulations.  Further, the orientational correlations decay much slower than translational correlations. This striking difference between translational and orientational correlations does not exist in the disordered liquid phase. In particular, our results in Fig.~\ref{fig:correlationFunctions}C for $T=312$\,K, which is above the order--disorder transition temperature, show short-range orientational correlations. 

The dislocations highlighted in Fig.~\ref{fig:correlationFunctions}A and the correlation functions in Fig.~\ref{fig:correlationFunctions}B, C and D are calculated for the plane consisting of tail-end particles of one monolayer. We have also studied the structure and the correlation functions for planes corresponding to each particle along the length of the tail. As could be anticipated~\cite{Seelig1974, Dill1980}, we observe that the number of dislocations decreases as we move upwards in the lipid chain. However, despite the disorder gradient, the translational and orientational correlations at every level along the lipid chains are consistent with the hexatic phase~(see \emph{SI}). The relative stability of the hexatic phase over a solid phase in this system can be interpreted in terms of the presence of thermally induced fluctuations in the midplane of the membrane \cite{Nelson1987}. These fluctuations can cause unbinding of dislocations, which destroys translational order. 
 
Taken together --- the finite concentration of diffusing 5\textendash 7 disclination pairs, the quasi-long-range orientational order, and the short-range translational order --- indicate that the ordered phase of the model lipid bilayer we study is a hexatic phase.

\subsection{Model dependence and relation to experiments} \ X-ray diffraction experiments have shown that an ordered hexatic phase exists in between the solid and disordered phases of lipid monolayers \cite{Kjaer1987}. In the case of lipid bilayers, the X-ray scattering data found a finite in-plane correlation length in the ordered phase, about 20\,nm~\cite{Smith1990}, but the experiments were unable to distinguish between a hexatic phase and a solid phase with finite sized domains.  Perhaps our prediction will motivate further experimental work to resolve this issue.

Bear in mind that while the MARTINI model is able to reproduce the order of the transition in DPPC bilayers, its ordered phase does not exhibit the experimentally observed tilt characteristic of this phase, and tilt can be induced in the MARTINI model by decreasing the size of tail particles relative to head particles~\cite{Marrink2005}.  Tilt can also be induced in a soft-sphere model by increasing repulsion between heads  \cite{Kranenburg2005}. Such features and the coupling between tilt and the order--disorder transition can be explored in future work.  

Also bear in mind that our analysis of the MARTINI model considers only reversible equilibrium behaviors.  For example, another feature of a fully hydrated phosphatidylcholine bilayer, observed in cooling protocols of multilamellar vesicles and planar bilayers, is the appearance of the so-called ripple phase, where the surface of the bilayer appears corrugated \cite{Tardieu1973}. The ripple structure is not observed in the equilibrium behavior of the MARTINI model, but it is worth studying whether this structure and others emerge out of equilibrium, as metastable states.  Moreover, both tilt as well as the rippled structure disappear upon addition of a small percentage of impurities such as cholesterol \cite{Hicks1987, Mills2009}, which is another behavior worthy of future study. 

Concerning our finding of first-order character in the liquid\textendash hexatic transition, experiments have found such behavior in five-layer thick smectic liquid-crystal films~\cite{Amador1989}.  Further, recent simulations indicate that the hexatic to liquid transition is in fact a first-order transition in two-dimensional systems of hard disks \cite{Bernard2011}.  Other work on two-dimensional systems shows that the order of the hexatic to liquid transition depends on the steepness of the repulsive interactions between particles \cite{Kapfer2015}. Specifically, it was found that for soft disks with repulsive potentials $r^{-n}$, the transition is continuous for $n < 6$ while it is first order for $n > 6$. Therefore, the order of the hexatic-to-liquid phase transition is a system-dependent property. Indeed, a different model for lipid bilayers with soft repulsive interactions \cite{Kranenburg2005, Venturoli2006} was found by Rodgers et al.\ to change continuously between ordered and disordered structures.   In that model, it is not even clear that a transition exists between distinct phases.

This system dependence is more than a theoretical curiosity.  Experiments can vary the presence and order of a transition by varying membrane composition. For compositions where the transition remains first order, pre-melting of the ordered phase and related phenomena become relevant.  In particular, large enough membrane-bound solutes, (e.g., transmembrane proteins) can form surfaces that disfavor the ordered phase even at thermodynamic conditions where the ordered phase is stable. At such conditions, a microscopic pre-melting layer with disordered membrane structure will then necessarily surround these solutes.  Several such neighboring solutes will then assemble to minimize line tension.  

This mechanism of self assembly is the subject of the subsequent paper \cite{Katira2015b}. The strength of the assembling force will depend upon the temperature, lateral pressure and the composition of the membrane, all of which affect the nature of the order--disorder transition.  Without its first-order character, solute-induced pre-melting layers will not exist, line tension will not exist, and this force of assembly will vanish.  We think the possibility of moving between regimes with differing transition order implies an important degree of complexity and correlated behavior yet to be explored.

\begin{materials}
We use the MARTINI coarse-grained force field \cite{Marrink2007} to model a lipid bilayer system in which four carbon atoms (or equivalent) are approximated as one coarse-grained bead. This model has an explicit solvent with approximately four water molecules scaled to one solvent particle. Dipalmitoyl phosphatidylcholine (DPPC) is chosen as the model lipid species. The MARTINI model uses the Lennard-Jones potential for non-bonded interactions. The cut-off for these interactions is 1.2\,nm. The GROMACS shifting function \cite{vanderspoel2002} is used in the range 0.99\textendash1.2  \,nm. Bond and angle energies are modeled as harmonic potentials with associated force constants 1250 \,kJmol$^{-1}$\,nm$^{-2}$ and 25\,kJmol$^{-1}$rad$^{-2}$ respectively. We use a time step of 30\,fs, within the recommended range of 20\textendash 40\,fs for this model. Simulations are performed using the GROMACS molecular dynamics package \cite{Pronk2013} in the ensemble with fixed numbers of particles, temperature, and stress tensor components for the system \cite{Zhang1995,Marrink2007,Rodgers2012}. Thermostats and barostats were used to control temperature and pressure, and checks were performed to assure that different thermostats and barostats yielded similar results \cite{Frenkel2001}. The compressibility is set to 3$\times$10$^{-5}$\,bar$^{-1}$. A lateral pressure of zero is maintained by choosing the diagonal components of the stress tensor to be 1\,bar. The electrostatic interactions are shifted to zero in the range 0\textendash 1.2 \,nm. A dielectric constant of 15 is used for screening of electrostatic interactions. 

The flat interface is stabilized by juxtaposing an ordered bilayer equilibrated at 285\,K and zero lateral pressure with a disordered bilayer equilibrated at the same conditions corresponding to the cooling and heating curves of the hysteresis loop in Fig.~\ref{fig:hysteresis}, respectively. The system thus constructed is equilibrated in the ensemble with fixed temperature, volume and numbers of particles. This ensemble allows for maintaining an area per lipid intermediate between the two phases, thus stabilizing the interface. 

The translational and orientational correlation functions for the ordered phase are calculated for a system containing 51,200 lipids solvated by 800,000 water molecules (1.4 million particles). This system is prepared by first equilibrating a smaller system containing 3,200 lipids to make a hydrated bilayer membrane of area roughly 26$\times$27\,nm$^2$.  A configuration from this equilibrated system is then periodically replicated to produce an initial configuration for the simulated system, which has a membrane area of roughly 107$\times$110\,nm$^2$. The system is then relaxed for about 11 microseconds resulting in a hexatic phase. 

To identify 5--7 disclination pairs, we use Shewchuk's code, Triangle~\cite{Shewchuk1, Shewchuk2}.

\end{materials}

%

%%%%%%%%
\begin{acknowledgments}
SV is currently supported by the University of Chicago. In the initial stages, he was supported by Director, Office of Science, Office of Basic Energy Sciences, Materials Sciences, and Engineering Division, of the U.S. Department of Energy under contract No.\ DE AC02-05CH11231.  KKM, DC, SK and BS are supported by that same DOE funding source, the latter two with FWP number SISGRKN. This research used resources of the National Energy Research Scientific Computing Center, a DOE Office of Science User Facility supported by the Office of Science of the U.S. Department of Energy under Contract No. DE-AC02-05CH11231 and resources of the Midway-RCC computing cluster at University of Chicago.
\end{acknowledgments}
%%%%%%%%
\bibliographystyle{pnas}
%\bibliography{AllBibs}

\begin{thebibliography}{10}

\bibitem{Katira2015b}
Katira, S, Mandadapu, K.~K, Vaikuntanathan, S, Smit, B,  \& Chandler, D.
\newblock (2015) The order\textendash disorder phase transition in lipid bilayers mediates a force for assembly of transmembrane proteins. {\em arXiv:1506.04298}.

\bibitem{Tardieu1973}
Tardieu, A, Luzzati, V,  \& Reman, F.~C.
\newblock (1973) Structure and polymorphism of the hydrocarbon chains of lipids: a study
of lecithin-water phases. {\em Journal of molecular biology} {\bf 75}:711--733.

\bibitem{Marrink2005}
Marrink, S.~J, Risselada, H.~J,  \& Mark, A.~E.
\newblock (2005) Simulation of gel phase formation and melting in lipid bilayers using a coarse grained model.  {\em Chemistry and physics of lipids} {\bf 135}:223--244.

\bibitem{Marrink2007}
Marrink, S.~J, Risselada, H.~J, Yefimov, S.~Y, Tieleman, D.~P,  \& de~Vries, A.~H.
\newblock (2007) The MARTINI force field: Coarse grained model for bimolecular simulations. {\em The Journal of Physical Chemistry B} {\bf 111}:7812--7824.

\bibitem{Rodgers2012b}
Rodgers, J.~M, Sorensen, J, de~Meyer, F. J.-M, Schiott, B,  \& Smit, B.
\newblock (2012) Understanding the phase behavior of coarse-grained model lipid bilayers
through computational calorimetry. {\em The Journal of Physical Chemistry B} {\bf 116}:1551--1569.

\bibitem{Buff1965}
Buff, F.~P, Lovett, R.~A,  \& Stillinger, F.~H.
\newblock (1965) Interfacial Density Profile for Fluids in the Critical Region. {\em Physical Review Letters} {\bf 15}:621--623.

\bibitem{Weeks1977}
Weeks, J.~D.
\newblock (1977) Structure and thermodynamics of the liquid--vapor interface. {\em The Journal of Chemical Physics} {\bf 67}:3106--3121.

\bibitem{Mabrey1976}
Mabrey, S \& Sturtevant, J.~M.
\newblock (1976) Investigation of phase transitions of lipids and lipid mixtures by
sensitivity differential scanning calorimetry. {\em Proceedings of the National Academy of Sciences} {\bf
 73}:3862--3866.

\bibitem{Nelson}
Nelson, D.~R.
\newblock (2002) {\em Defects and Geometry in Condensed Matter Physics}.
\newblock (Cambridge University Press).

\bibitem{Halperin1978}
Halperin, B.~I \& Nelson, D.~R.
\newblock (1978) Theory of two-dimensional melting. {\em Physical Review Letters} {\bf 41}:121--124.

\bibitem{Marcus1997}
Marcus, A.~H \& Rice, S.~A.
\newblock (1997) Phase transitions in a confined quasi-two-dimensional colloid
suspension. {\em Physical Review E} {\bf 55}:637.

\bibitem{Machta2012}
Machta, B.~B, Veatch, S.~L,  \& Sethna, J.~P.
\newblock (2012)  Critical Casimir forces in cellular membranes. {\em Physical Review Letters} {\bf 109}:138101.

\bibitem{Berger1997}
Berger, O, Edholm, O,  \& J{\"a}hnig, F.
\newblock (1997) Molecular dynamics simulations of a fluid bilayer of
dipalmitoylphosphatidylcholine at full hydration, constant pressure, and
constant temperature. {\em Biophysical Journal} {\bf 72}:2002.

\bibitem{Klauda2010}
Klauda, J.~B, Venable, R.~M, Freites, J.~A, O¿Connor, J.~W, Tobias, D.~J,
  Mondragon-Ramirez, C, Vorobyov, I, MacKerell~Jr, A.~D,  \& Pastor, R.~W.
\newblock (2010) Update of the CHARMM all-atom additive force field for lipids:
validation on six lipid types. {\em The Journal of Physical Chemistry B} {\bf 114}:7830--7843.

\bibitem{Cooke2005}
Cooke, I.~R \& Deserno, M.
\newblock (2005) Solvent-free model for self-assembling fluid bilayer membranes:
stabilization of the fluid phase based on broad attractive tail potentials. {\em The Journal of Chemical Physics} {\bf 123}:224710.

\bibitem{Venturoli2005}
Venturoli, M, Smit, B,  \& Sperotto, M.~M.
\newblock (2005) Simulation studies of protein-induced bilayer deformations, and
lipid-induced protein tilting, on a mesoscopic model for lipid bilayers with
embedded proteins. {\em Biophysical Journal} {\bf 88}:1778--1798.

\bibitem{Lenz2005}
Lenz, O \& Schmid, F.
\newblock (2005) A simple computer model for liquid lipid bilayers. {\em Journal of Molecular Liquids} {\bf 117}:147--152.

\bibitem{Ingolfsson2014}
Ing{\'o}lfsson, H.~I, Melo, M.~N, van Eerden, F.~J, Arnarez, C, Lopez, C.~A,
  Wassenaar, T.~A, Periole, X, De~Vries, A.~H, Tieleman, D.~P,  \& Marrink,
  S.~J.
\newblock (2014) Lipid organization of the plasma membrane. {\em Journal of the American Chemical Society} {\bf 136}:14554--14559.

\bibitem{Risselada2008}
Risselada, H.~J \& Marrink, S.~J.
\newblock (2008) The molecular face of lipid rafts in model membranes. {\em Proceedings of the National Academy of Sciences} {\bf
  105}:17367--17372.

\bibitem{Schafer2011}
Sch{\"a}fer, L.~V, de~Jong, D.~H, Holt, A, Rzepiela, A.~J, de~Vries, A.~H,
  Poolman, B, Killian, J.~A,  \& Marrink, S.~J.
\newblock (2011) Lipid packing drives the segregation of transmembrane helices into
disordered lipid domains in model membranes. {\em Proceedings of the National Academy of Sciences} {\bf
  108}:1343--1348.

\bibitem{Baoukina2012}
Baoukina, S, Marrink, S.~J,  \& Tieleman, D.~P.
\newblock (2012) Molecular structure of membrane tethers. {\em Biophysical Journal} {\bf 102}:1866--1871.

\bibitem{Willard2010}
Willard, A.~P \& Chandler, D.
\newblock (2010) Instantaneous liquid interfaces. {\em The Journal of Physical Chemistry B} {\bf 114}:1954--1958.

\bibitem{Limmer2014}
Limmer, D.~T \& Chandler, D.
\newblock (2014) Premelting, fluctuations, and coarse-graining of water-ice interfaces. {\em The Journal of Chemical Physics} {\bf 141}:18C505.

\bibitem{Safran1994}
Safran, S.~A.
\newblock (1994) {\em Statistical thermodynamics of surfaces and interfaces}.
\newblock (Addison-Wesley, New York).

\bibitem{Nelson1982}
Nelson, D.~R, Rubinstein, M,  \& Spaepen, F.
\newblock (1982) Order in two-dimensional binary random arrays. {\em Philosophical Magazine A} {\bf 46}:105--126.

\bibitem{Kosterlitz1973}
Kosterlitz, J.~M \& Thouless, D.~J.
\newblock (1973) Ordering, metastability and phase transitions in two-dimensional systems. {\em Journal of Physics C: Solid State Physics} {\bf 6}:1181--1203.

\bibitem{Kosterlitz1978}
Kosterlitz, J.~M \& Thouless, D.~J.
\newblock (1978) Two-dimensional physics. {\em Progress in Low Temperature Physics} {\bf 7}:373--433.

\bibitem{Nelson1979}
Nelson, D.~R \& Halperin, B.~I.
\newblock (1979) Dislocation-mediated melting in two dimensions. {\em Physical Review B} {\bf 19}:2457--2484.

\bibitem{Young1979}
Young, A.~P.
\newblock (1979) Melting and the vector coulomb gas in two dimensions. {\em Physical Review B} {\bf 19}:1855--1866.

\bibitem{Birgeneau1978}
Birgeneau, R.~J \& Lister, J.~D.
\newblock (1978) Bond orientational order model for smectic B liquid crystals. {\em Journal de Physique Lettres} {\bf 39}:399--402.

\bibitem{Pindak1981}
Pindak, R, Moncton, D.~E, Davey, S.~C, \& Goodby, J.~W.
\newblock (1981) X-ray observation of a stacked hexatic liquid-crystal B phase. {\em Physical Review Letters} {\bf 46}:1135.

\bibitem{Kjaer1987}
Kjaer, K, Als-Nielsen, J, Helm, C.~A, Laxhuber, L.~A,  \& M{\"o}hwald, H.
\newblock (1987) Ordering in lipid monolayers studied by synchrotron X-ray diffraction
and fluorescence microscopy. {\em Physical Review Letters} {\bf 58}:2224.

\bibitem{Brock1989}
Brock, J.~D, Noh, D.~Y, McClain, B.~R, Lister, J.~D, Birgeneau, R.~J, Aharony, A, Horn, P.~M, \& Liang, J.~C.
\newblock (1989) Hexatic ordering in freely suspended liquid crystal films. {\em Zeitschrift f{\"u}r Physik B Condensed Matter} {\bf 74}:197--213.

\bibitem{Zangi1998}
Zangi, R \& Rice, S.~A.
\newblock (1998) Phase transitions in a quasi-two-dimensional system. {\em Physical Review E} {\bf 58}:7529.

\bibitem{Bernard2011}
Bernard, E.~P \& Krauth, W.
\newblock (2011) Two-step melting in two dimensions: First-order liquid-hexatic transition. {\em Physical Review Letters} {\bf 107}:155704.

\bibitem{Strandburg1988}
Strandburg, K.~J.
\newblock (1988) Two-dimensional melting. {\em Reviews of Modern Physics} {\bf 60}:161.

\bibitem{Seelig1974}
Seelig, A \& Seelig, J.
\newblock (1974) Dynamic structure of fatty acyl chains in a phospholipid bilayer
measured by deuterium magnetic resonance. {\em Biochemistry} {\bf 13}:4839--4845.

\bibitem{Dill1980}
Dill, K.~A \& Flory, P.~J.
\newblock (1980) Interphases of chain molecules: Monolayers and lipid bilayer
membranes. {\em Proceedings of the National Academy of Sciences} {\bf
  77}:3115--3119.
  
\bibitem{Nelson1987}
Nelson, D.~R \& Peliti, L.
\newblock (1987) Fluctuations in membranes with crystalline and hexatic order. {\em Journal de Physique} {\bf 48}:1085--1092.

\bibitem{Smith1990}
Smith, G, Sirota, E, Safinya, C, Plano, R,  \& Clark, N.
\newblock (1990) X-ray structural studies of freely suspended hydrated DMPC multimembrane films. {\em The Journal of Chemical Physics} {\bf 92}:4519--4529.

\bibitem{Kranenburg2005}
Kranenburg, M \& Smit, B.
\newblock (2005) Phase behavior of model lipid bilayers. {\em The Journal of Physical Chemistry B} {\bf 109}:6553--6563.

\bibitem{Hicks1987}
Hicks, A, Dinda, M,  \& Singer, M.~A.
\newblock (1987) The ripple phase of phosphatidylcholines: effect of chain length and
cholesterol. {\em Biochimica et Biophysica Acta (BBA)-Biomembranes} {\bf
  903}:177--185.

\bibitem{Mills2009}
Mills, T.~T, Huang, J, Feigenson, G.~W,  \& Nagle, J.~F.
\newblock (2009) Effects of cholesterol and unsaturated DOPC lipid on chain packing of
saturated gel-phase DPPC bilayers. {\em General Physiology and Biophysics} {\bf 28}:126.

\bibitem{Amador1989}
Amador, S, Pershan, P.~S, Stragier, H, Swanson, B.~D, Tweet, D.~J, Sorensen, L.~B, Sirota, E.~B, Ice, G.~E, \& Habenschuss, A.
\newblock (1989) Synchrotron studies of the first-order melting transitions of hexatic monolayers and multilayers in freely suspended liquid-crystal films. {\em Physical Review A} {\bf 39}:2703.

\bibitem{Kapfer2015}
Kapfer, S.~C \& Krauth, W.
\newblock (2015) Two-dimensional melting: from liquid-hexatic coexistence to continuous
transitions. {\em Physical Review Letters} {\bf 114}:035702.

\bibitem{Venturoli2006}
Venturoli, M \& Smit, B.
\newblock (1999) Simulating the self-assembly of model membranes. {\em Physical Chemistry Communications} {\bf 2}:45--49.
	
\bibitem{vanderspoel2002}
van~der Spoel, D, van Buuren, A.~R, Apol, M.~E.~F, Meulenhoff, P.~J, Tieleman, D.~P, Sijbers,
  A.~L.~T.~M, Hess, B, Feenstra, K.~A, Lindahl, E, van Drunen, R,  \& Berendsen, H.~J.~C.
\newblock (2002) {\em http://www.gromacs.org/.}

\bibitem{Pronk2013}
Pronk, S, Pail, S, Schulz, R, Larsson, P, Bjelkmar, P, Apostolov, R, Shirts, M.~R,
  Smith, J.~C, Kasson, P.~M, van~der Spoel, D, Hess, B,  \& Lindahl, E.
\newblock (2013) GROMACS 4.5: A high-throughput and highly parallel open source molecular simulation toolkit. {\em Bioinformatics} {\bf 29}:845--854.

\bibitem{Zhang1995}
Zhang, Y, Feller, F.~E, Brooks, B.~R,  \& Pastor, R.~W.
\newblock (1995) Computer simulation of liquid/liquid interfaces. I. Theory and application to octane/water. {\em The Journal of Chemical Physics} {\bf 103}:10252--10266.

\bibitem{Rodgers2012}
Rodgers, J.~M \& Smit, B.
\newblock (2012) On the equivalence of schemes for simulation bilayers at constant surface tension. {\em Journal of Chemical Theory and Computation} {\bf 8}:404--417.

\bibitem{Frenkel2001}
Frenkel, D \& Smit, B. 
\newblock (2001) Understanding molecular simulation: from algorithms to applications. {\em Academic Press}.

\bibitem{Shewchuk1}
Shewchuk, J.~R.
\newblock (1996) Triangle:  Engineering a 2D Quality Mesh Generator and Delaunay Triangulator. 
{\em Applied Computational Geometry:  Towards Geometric Engineering, Lecture Notes in Computer Science, Eds. Lin, M.~C \& Manocha, D } {\bf 1148}:203--222. 

 
\bibitem{Shewchuk2}
Shewchuk, J.~R.
\newblock (2002) Delaunay Refinement Algorithms for Triangular Mesh Generation. {\em Computational Geometry: Theory and Applications} {\bf 22}:21--74. 

\end{thebibliography}

\end{article}

\clearpage

%\title{Supporting Information} 

%\author{K. Hashimoto et al.}

%\contributor{Submitted to Proceedings of the National Academy of Sciencesof the United States of America}

%\maketitle

\renewcommand{\thefigure}{S\arabic{figure}}

%\begin{article}
%\tableofcontents
%\listoffigures

\noindent
{\huge\bf Supporting Information} \\
\\
{\Large\bf for ``... is a first-order hexatic to liquid phase transition''} 

\bigskip
\noindent
{\large\bf by Shachi Katira, Kranthi K. Mandadapu, Suriyanarayanan Vaikuntanathan, Berend Smit and David Chandler}
 %{\normalsize\\ \vspace{0.1cm} *Equal contribution} 
  
\setcounter{figure}{0} \renewcommand{\thefigure}{S\arabic{figure}}
\setcounter{equation}{0} \renewcommand{\theequation}{S\arabic{equation}}

\vspace{0.25 in}

In this Supporting Information, we provide evidence that the different planes containing the different beads along the length of the chain also have short-range translational order and quasi-long-range orientational order, thus indicating that the entire hydrophobic part of the bilayer is in the hexatic phase.  We also discuss the relaxation of the bilayer system into the hexatic phase and the corresponding relaxation of the translational, pair and orientational correlation functions for the million-particle system. 
                 
\section{Along the lipid chain}

\begin{figure*}[h!]
\centering
\includegraphics[scale=0.185]{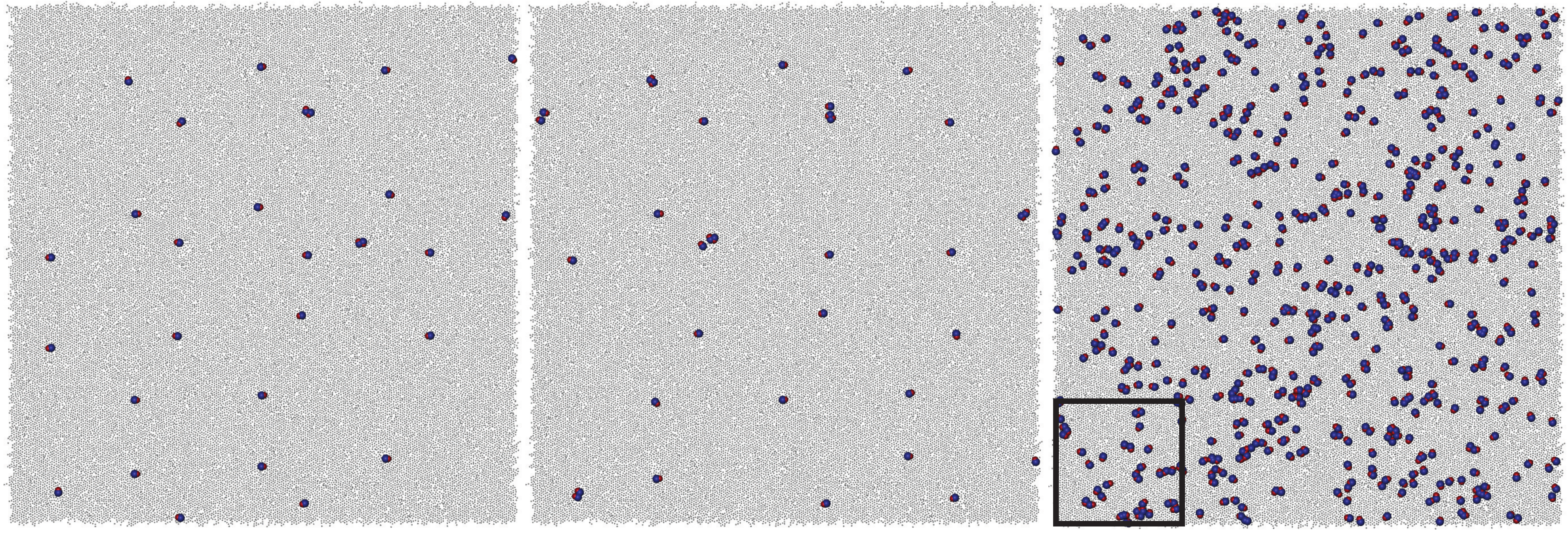}
\caption{\label{fig:57PairsSI} 5\textendash 7 disclination pairs for planes containing C2 (left), C3 (center) and C4 (right) particles in a 107$\times$110\,nm$^2$ bilayer in the ordered phase at $T$=294\,K. The tail-end particles of one monolayer are shown in gray. Enlarged red and blue circles render particles with seven and five neighbors, respectively. For each plane, there exist unbound dislocations as expected for a hexatic phase, and the number of dislocations increases towards the tail-end.  The box drawn in for the C4 panel shows the region illustrated in Panel A of Fig. 4 in the main text. }
\end{figure*}

\begin{figure*}[h!]
\centering
\includegraphics[scale=0.42]{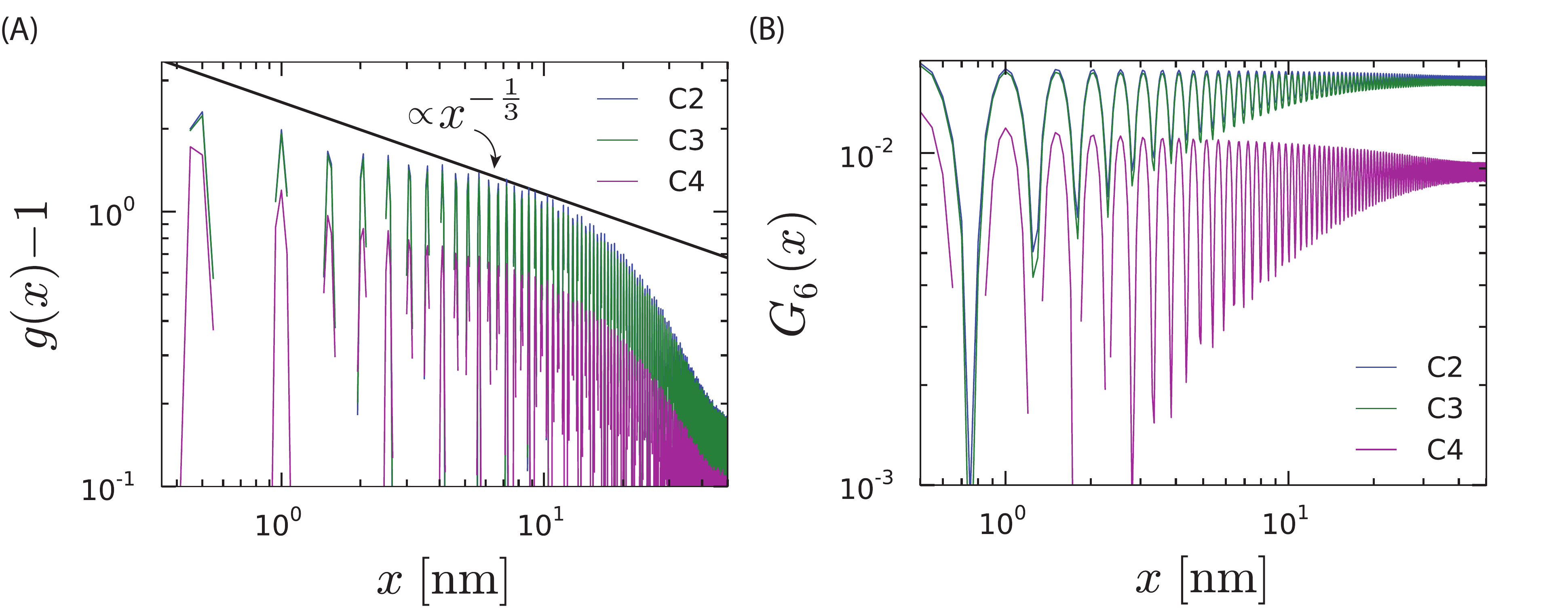}
\caption{\label{fig:C234}Pair correlations in the hexatic phase: (A)  $g(x)$ for tail particles C2, C3 and the tail-end C4, showing decay faster than $x^{-1/3}$. (B) $G_6(x)$ showing quasi-long-range orientational order for all three planes of particles. }
\end{figure*}

In addition to the tail-end particles (C4), we  calculate pair correlation functions for non-tail-end particles C2 and C3 on the lipid chains. The planes containing these particles show fewer unbound dislocations compared to the tail-end particles C4 as shown in Fig.~\ref{fig:57PairsSI}. The amount of disorder therefore decreases higher along the lipid chains, consistent with the disorder gradient observed in experiments \cite{Seelig1974} and explained by theory \cite{Dill1980}. Despite the gradient in disorder, the pair correlation functions, $g(x)$, seen in Fig.~\ref{fig:C234}A appear to decay exponentially for all three planes of particles. The orientational correlation functions also exhibit quasi-long-range behavior Fig.~\ref{fig:C234}B.  Moreover, the dislocations in the C2 and C3 planes also diffuse freely in the time scale of the simulation, albeit slower than those in the C4 plane. These data show that all three planes of particles are hexatic. 

\section{Relaxation}

\begin{figure*}[h!]
\centering
\includegraphics[scale=0.42]{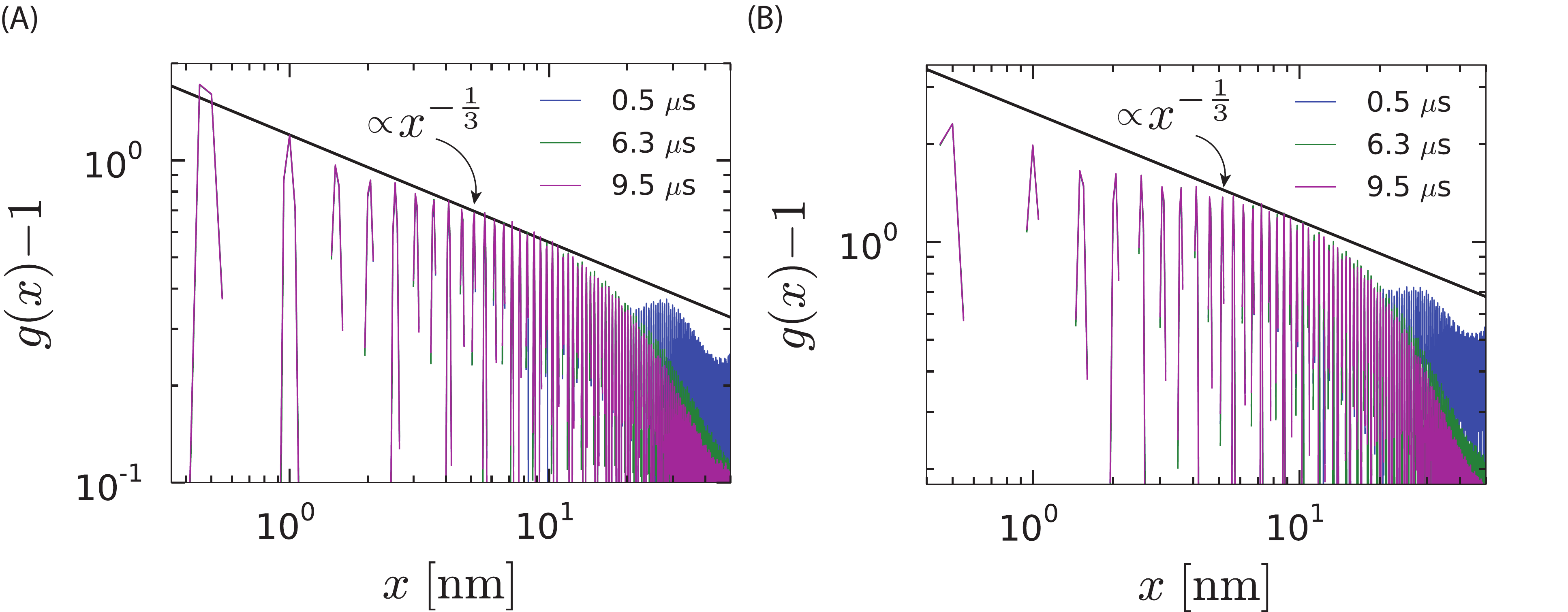}
\caption{\label{fig:timegx} Translational pair correlation function during relaxation to the hexatic phase, starting from initial non-equilibrium state.  (A) $g(x)-1$ for tail particles C4 at three different times.  (B) $g(x)-1$ for tail particles C2 at three different times.}
\end{figure*}

In Fig.~\ref{fig:timegx}, we show the time evolution of the pair correlation function for the most ordered C2 and the least ordered C4 planes. The correlations observed at 0.5 microseconds reflect the preparation of the system by periodically replicating a small ordered patch several times in the $x$ and $y$ directions.  Specifically, notice the broad peaks at those early times, as the initial configuration is a lattice, effectively a crystal, with large unit cells.  As time progresses, the lattice dissipates, positions become increasingly uncorrelated, and exponential decay emerges for $g(x)-1$ in both C2 and C4 planes.

\begin{figure*}[t!]
\centering
\includegraphics[scale=0.42]{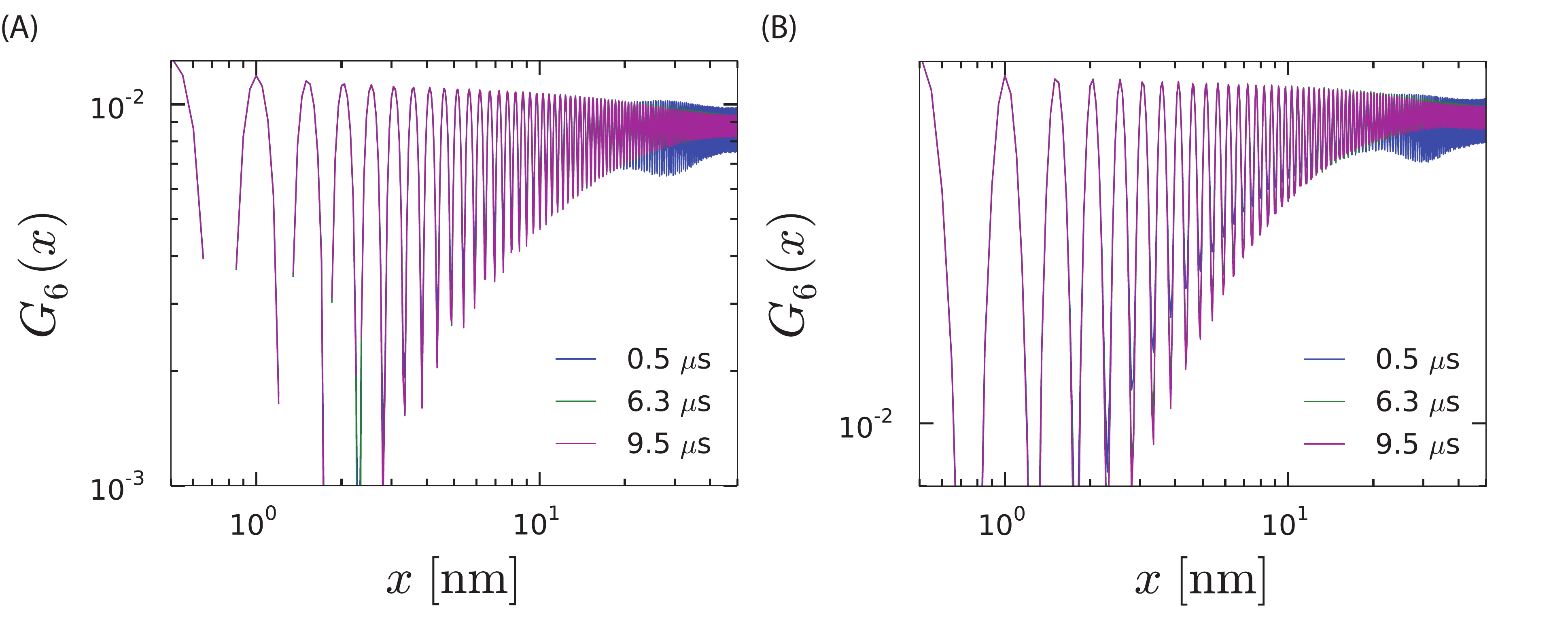}
\caption{\label{fig:timeg6} Orientational pair correlation function during relaxation to the hexatic phase, starting from initial non-equilibrium state.  (A) $G_6(x)$ for tail particles C4 at three different times.  (B) $G_6(x)$ for tail particles C2 at three different times.}
\end{figure*}

In Fig.~\ref{fig:timeg6}, we show the time evolution of the orientational correlation function $G_6(\mathbf{r})$ for the C2 and C4 planes of particles. The long-range correlations introduced by periodically replicating the system gradually diminish and the system shows a quasi-long-range behavior at 6.3\,$\mu$s.

\end{document}